\documentclass[%
 reprint,
 amsmath,
 amssymb,
 aps,
 prd,
]{revtex4-1}

\usepackage{graphicx}
\usepackage{dcolumn}
\usepackage{bm}
\usepackage{hyperref}
\usepackage{natbib}
\hypersetup{
    colorlinks=true,
    linkcolor=blue,
    filecolor=blue,      
    urlcolor=blue,
    citecolor=blue,
    }

\begin{document}

\preprint{APS/123-QED}

\title{\textbf{From eccentric binaries to nonstationary gravitational wave backgrounds} 
}%

\author{Mikel Falxa}%
 \email{Contact author: mikel.falxa@unimib.it}
\affiliation{%
 Authors' affiliations\\
  Dipartimento di Fisica “G. Occhialini", Università degli Studi di Milano-Bicocca, Piazza della Scienza 3, I-20126 Milano, Italy.
}%

\author{Hippolyte Quelquejay Leclere}%
\affiliation{%
 Authors' affiliations\\
  Université Paris Cité, CNRS, Astroparticule et Cosmologie, F-75013 Paris, France
}%

\author{Alberto Sesana}%
\affiliation{%
 Authors' affiliations\\
  Dipartimento di Fisica “G. Occhialini", Università degli Studi di Milano-Bicocca, Piazza della Scienza 3, I-20126 Milano, Italy}
\affiliation{INFN - Sezione di Milano-Bicocca,
Piazza della Scienza 3, 20126 Milano, Italy}
\affiliation{INAF - Osservatorio Astronomico di Brera,
    Via Brera 20, 20121 Milano, Italy
}%

\date{\today}

\begin{abstract}
A large population of binary systems in the Universe emitting gravitational waves (GW) would produce a stochastic noise, known as the gravitational wave background (GWB). The properties of the GWB directly depend on the attributes of its constituents. If the binary systems are in eccentric orbits, it is well established that the GW power they radiate strongly depends on their instantaneous orbital phase. Consequently, their power spectrum varies over time, and the resulting GWB can appear nonstationary. In this work, we estimate the amplitude of time-dependent fluctuations in the GWB power spectrum as a function of the eccentricity of the binaries. Specifically, we focus on the GWB produced by a population of supermassive black hole binaries (SMBHB) that should be observable by pulsar timing arrays (PTA). We show that a large population of homogeneously distributed equal SMBHBs produces nonstationary features that are undetectable by current PTA datasets. However, using more realistic and astrophysically motivated populations of SMBHBs, we show that the nonstationarity might become very large and detectable, especially in the case of more massive and eccentric populations. In particular, when one binary is slightly brighter than the GWB, we demonstrate that time fluctuations can become significant. This is also true for individual binary systems with a low signal-to-noise ratio (SNR) relative to the GWB (SNR $\approx$ 1), which standard data analysis methods would struggle to detect. The detection of nonstationary features in the GWB could indicate the presence of some relatively bright GW sources in eccentric orbits, offering new insights into the origins of the signal. This will be particularly relevant for PTAs using observations from the next generation of radio telescopes, as they will offer high-precision measurements of GWB properties and increased sensitivity to time-dependent fluctuations.
\end{abstract}

\maketitle


\section{Introduction}

The gravitational wave (GW) energy radiated by an eccentric binary system depends on its instantaneous phase and frequency. When the two orbiting objects are at periapsis, the acceleration they experience is higher, and thus the emitted gravitational energy is stronger. The higher the eccentricity of the system, the greater its average gravitational luminosity. This was first demonstrated in \cite{PM}, where a full derivation of the GW radiation from eccentric binaries was proposed at leading post-Newtonian order. For high eccentricities, the emitted energy can vary widely within one orbital period of the binary. It is therefore relevant to consider whether this variation can produce nonstationary signals, particularly whether a population of eccentric binaries will generate stochastic noise with nonstationary features.

In this work, we aim to derive the expected amplitude of time-dependent fluctuations in a stochastic gravitational wave background (GWB) generated by a population of binary systems in eccentric orbits (e.g., binary black holes, galactic binaries, extreme mass ratio inspirals, etc.). Such GWBs can be detected by Pulsar Timing Arrays (PTAs) \cite{PM} or LISA \cite{lisa_red_book}. Recently, PTA collaborations have found evidence of a GWB-like signal that could be produced by a population of supermassive black hole binaries (SMBHBs) \cite{nanograv_gwb, wm3, ppta_gwb}. However, there is still no definitive interpretation of the observed signal, and other sources could be at the origin of that signal \cite{epta_interpretation}. Searching for nonstationary features could provide more information about its origin, particularly if it results from a superposition of SMBHBs in eccentric orbits.

We organize this work as follows: in \autoref{sec:GW_ecc}, we present the physics of GWs radiated by eccentric binary systems. In \autoref{sec:GWB_ecc} and \autoref{sec:astro_effect}, we introduce a quantity that measures the amplitude of time-dependent fluctuations for nonstationary noise and apply it to a population of eccentric binaries to estimate the order of magnitude of the expected fluctuations. Then, \autoref{sec:time_fluct_cgw} explores the case of relatively loud individual binaries that may produce strong time-dependent fluctuations. Finally, in \autoref{sec:pta_sensitivity}, we qualitatively estimate PTA sensitivity to time fluctuations given their noise properties.

\section{GW from eccentric binaries}
\label{sec:GW_ecc}

\subsection{Radiated power}

The power radiated as GWs by a binary system can be estimated if we know all the components of its quadrupole moment $I^{ij}$. In \cite{PM}, the GW luminosity averaged over all directions is given by:

\begin{equation}
    L_{GW} = \frac{G}{5c^5} \left(\dddot{I}^{ij} \dddot{I}_{ij} - \frac{1}{3} (\dddot{I}^k_k)^2 \right),
\label{eq:quadrupole_formula}
\end{equation}
where the dots represent time derivatives.

\subsubsection{Power as a function of true anomaly}
\label{sec:power_psi}

Peters and Mathews \cite{PM} showed that the average luminosity over one orbital period $1/f_p$ of an eccentric binary with chirp mass $\mathcal{M}$ and eccentricity $e$ is given by

\begin{equation}
    \langle L_{GW} \rangle = P_0 F(e),
\end{equation}
with $\langle x \rangle$ denoting the average over one orbital period and $P_0 = (32/5c^5) G^{7/3} (\mathcal{M} 2\pi f_p )^{10/3}$ the power emitted if the binary was circular. The luminosity depends greatly on the eccentricity of the binary through the amplification factor $F(e)$, given by

\begin{equation}
    F(e) = \frac{\left( 1 + \frac{73}{24} e^2 + \frac{37}{96} e^4\right)}{(1 - e^2)^{7/2}}.
\end{equation}

In this article, we quantify the fluctuations of the GW luminosity within one orbital period of the binary by calculating the variance of $L_{GW}$. We find that

\begin{equation}
    \langle L_{GW}^2 \rangle - \langle L_{GW} \rangle^2 = P_0 ^2 [G^2(e) - F^2(e)],
\label{eq:total_variance}
\end{equation}
where the factor $G(e)$ is given by
\begin{equation}
    G^2(e) = \frac{1 + \frac{271}{12} e^2 + \frac{10155}{128} e^4 + \frac{50966}{768} e^6 + \frac{76735}{6144} e^8 + \frac{1027}{4096} e^{10}}{(1 - e^2)^{17/2}}.
\end{equation}

\begin{figure}
   \centering
   \includegraphics[width=0.5\textwidth]{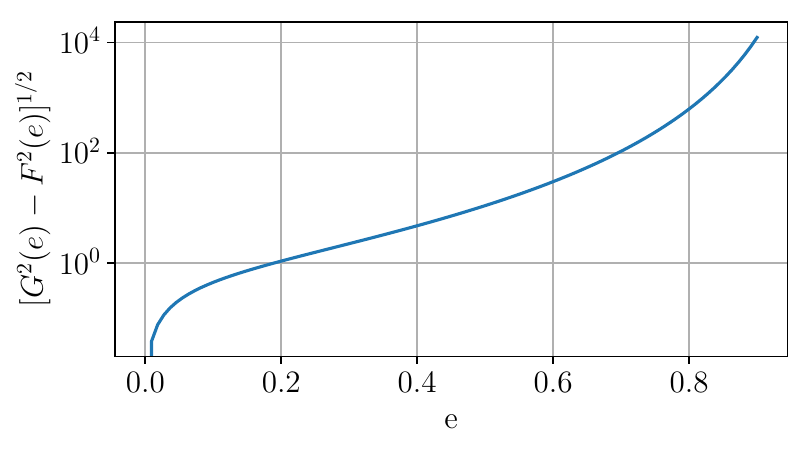}
   \caption{Variance of $L_{GW}/P_0$ as a function of the eccentricity of the source $e$.}
\label{fig:sigma_e}
\end{figure}

Considering a population of independent binaries, if we randomly choose a binary within that population at an instant $t$, its orbital phase (or true anomaly) will be random. The variance in \autoref{eq:total_variance} quantifies how much the measured GW luminosity of that binary can fluctuate around the expected average luminosity $\langle L_{GW} \rangle$. In fact, we could also calculate higher moments than the variance to completely portray the fluctuations. In this work, we decided to focus on the contribution of the second moment. In \autoref{fig:sigma_e}, we see that for circular binaries with $e=0$, the variance goes to 0.

\subsubsection{Fourier analysis of Keplerian orbit}

The Fourier analysis of the Keplerian orbit provides us with a harmonic decomposition of the quadrupole moment, as defined in \autoref{eq:quadrupole_formula}

\begin{equation}
    I^{ij} = \sum_n I^{ij}_n.
\label{eq:fourier_quadrupole}
\end{equation}

This decomposition allows us to write the total luminosity as a sum of harmonic contributions $P_n$

\begin{equation}
    L_{GW} = \sum_n P_n.
\end{equation}

In the following, we estimate the variance-covariance of the $P_n$. More detailed derivations are provided in \autoref{app:fourier_kepler}.

\subsubsection{First moment}

Consider a binary with orbital period $T_p$ and orbital frequency $f_p = 1/T_p$. Following \cite{PM}, we use $\Phi(t) \approx 2\pi f_p (t - t_0)$ and neglect the frequency evolution due to the GW radiation of the binary. Using \autoref{eq:fourier_quadrupole}, we can split the luminosity on a basis $f_n(t, e)$ constructed from \autoref{eq:fourier_basis_lgw} as

\begin{equation}
    P_n = \langle P_n \rangle f_n (t, e),
\end{equation}
with $\langle f_n (t, e) \rangle = 1$, and the brackets $\langle x \rangle$ denote the average over one orbital period. In \cite{PM}, it is shown that

\begin{equation}
\begin{aligned}
    \langle L_{GW} \rangle & = \sum_n \langle P_n \rangle \\
    & = P_0 \sum_n g(n, e),
\end{aligned}
\end{equation}
where $g(n, e)$ gives the average contribution of the $n$-th harmonic to the GW spectrum for a binary with eccentricity $e$. This result is rederived from the Fourier decomposition of the quadrupole moment in \autoref{app:fourier_kepler}.

\subsubsection{Second moment}

Now that we have an expression for $P_n$, it is straightforward to calculate the covariance between two different components $P_n$ and $P_m$. We apply the formula for the covariance $cov \{P_n, P_m \}$

\begin{equation}
\begin{aligned}
    \langle P_n P_m \rangle - \langle P_n \rangle \langle P_m \rangle & = P_0 ^2 g(n, e) g(m, e) \\
    & \times [\left \langle f_n (t, e) f_m(t, e) \right \rangle - 1 ].
\end{aligned}
\label{eq:cov_timefrequency}
\end{equation}

\begin{figure}
   \centering
   \includegraphics[width=0.5\textwidth]{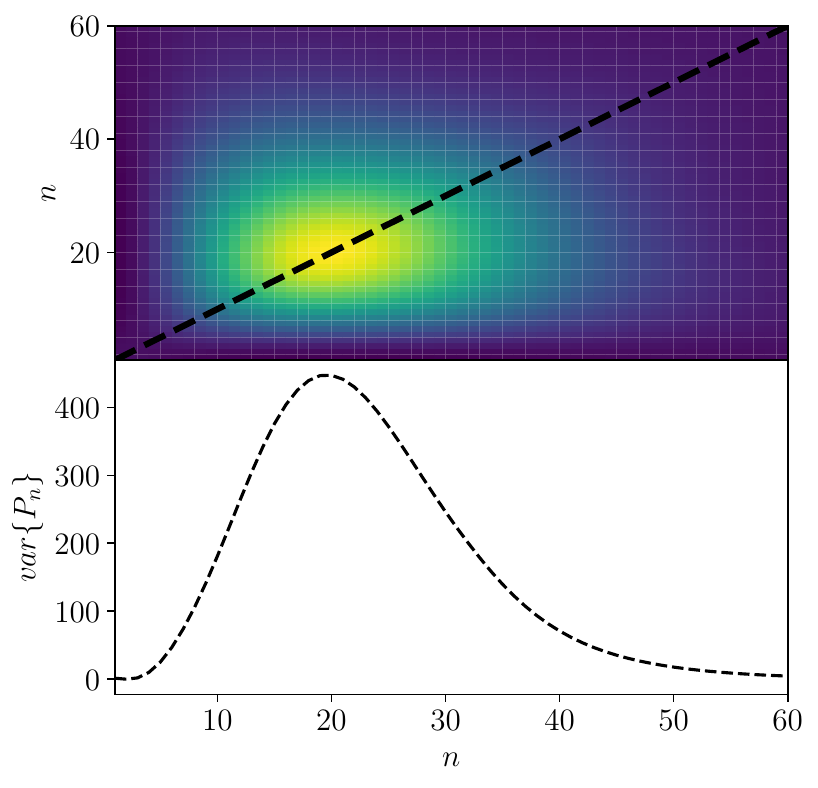}
   \caption{Covariance and variance for the first 60 harmonics of a GW spectrum radiated by a binary system with eccentricity $e=0.8$; (upper panel) covariance matrix $\langle P_n P_m \rangle - \langle P_n \rangle \langle P_m \rangle$, (lower panel) variance of the $P_n$, i.e., diagonal terms of the covariance matrix.}
\label{fig:covariance_variance}
\end{figure}

Note that for $n=m$, $cov \{P_n, P_m \}$ is the variance of $P_n$, $var \{P_n\}$. The $f_n(t,e)$ do not form an orthogonal basis, so $\langle f_n (t, e) f_m(t, e) \rangle \neq 0$. The covariance between harmonics generated by the same binary system is the result of its continuously changing velocity throughout the orbit. The contribution of consecutive higher harmonics increases "in phase" as the binary accelerates. We can see in \autoref{fig:covariance_variance} that neighboring harmonics are highly correlated. This correlation would be destroyed if each harmonic had different phases (e.g., when we compute the correlations between harmonics generated by two different binaries).

The covariance $\langle P_n P_m \rangle - \langle P_n \rangle \langle P_m \rangle$ can be easily obtained numerically. We can verify the validity of this expression by equating \autoref{eq:cov_timefrequency} and \autoref{eq:total_variance} as

\begin{equation}
    \langle L_{GW}^2 \rangle - \langle L_{GW} \rangle^2 = \sum_{n, m} \langle P_n P_m \rangle - \langle P_n \rangle \langle P_m \rangle.
\end{equation}

This equality is verified in the appendix (see \autoref{fig:tot_cov_error}) where we show that the relative error between the two sides is $\epsilon \approx 10^{-14}$.

In the rest of this article, we will consider the $P_n$ as random variables. The variance of each $P_n$ will characterize the fluctuations of the GW radiation within one orbital period of the binary. The spectral energy density is defined as \cite{Enoki_2007}

\begin{equation}
    \frac{d L_{GW}}{df} = \sum_n P_n \delta (f - nf_p).
\end{equation}

The trick that is used here is to account for the time-variability of the signal by adding random properties to the $P_n$. The dependence of the GW luminosity on the true anomaly is modeled by the randomness of the $P_n$. Essentially, the true anomaly is a random quantity, and the time average operation $\langle x \rangle$ is mapped to the average with respect to the true anomaly (orbital phase).

\subsection{Time fluctuations}

We define the time fluctuation $\Delta \Omega$ as the ratio between the standard deviation and the mean of the time-dependent spectrum $S(f, t)$ as in \cite{ns_fluct}. In order to be detectable, the fluctuations of a GWB should be at least at a level comparable to the uncertainties on the estimate of its spectrum amplitude (see \autoref{sec:pta_sensitivity}). This uncertainty depends on the noise properties and configuration of the considered detector \cite{forecasting}. The brackets now denote the average over time $1/f$.

\begin{equation}
    \Delta \Omega^2 (f) = \frac{|\langle S(f, t)^2\rangle - \langle S(f, t) \rangle^2|}{\langle S(f, t) \rangle^2},
\label{eq:fluct}
\end{equation}

where the numerator is the variance of the spectrum $S(f, t)$ and the denominator the mean spectrum squared. Considering previous works \cite{falxa_ns}, if we write $S(f, t) = g^2(f, t) S_0(f)$ with $\langle g^2(f, t) \rangle = 1$, we have

\begin{equation}
    \Delta \Omega ^2 (f) = |\langle g^4 (f, t) \rangle - 1|
\end{equation}
which can be measured in real data using analysis pipelines. Since the GW signal generated by eccentric binaries can exhibit correlations between different frequencies, it is interesting to define the two-frequency correlator of the measured fluctuations averaged over the total time of observation $T$

\begin{equation}
    \Delta \Omega^2 (f, f') = \frac{\langle S(f, t) S(f', t) \rangle_T - \langle S(f, t) \rangle_T \langle S(f', t) \rangle_T}{\langle S(f, t) \rangle_T \langle S(f', t) \rangle_T},
\label{eq:corr_fluct}
\end{equation}
that can provide additional information on the nature of the signal. In particular, it can be useful when dealing with single sources where correlations are strong (see \autoref{sec:time_fluct_cgw}) to estimate the total fluctuations.

\section{Gravitational wave background from eccentric binaries}
\label{sec:GWB_ecc}

\subsection{Gravitational wave backgrounds}

\subsubsection{GWB spectrum}

A population of binary systems individually emitting GW signals results in a stochastic GW background noise \cite{phinney}. The properties of this gravitational wave background (GWB) depend on the characteristics of its constituents \cite{Enoki_2007, Bonetti_2024, Chen_2019}. For a population of GW-driven binary systems, we deduce the power spectral density (PSD) of the GWB by invoking ergodicity \cite{Enoki_2007, Christensen_2018}: we can equate the average energy emitted by one GW-driven binary during its entire lifetime with the ensemble average, at time $t$, of a population of binaries at different stages of their lives. For circular binaries, this calculation is straightforward since binary systems emit monochromatic GWs at a steady rate, and one only needs to know how the population of binaries is distributed in orbital frequency. However, when considering eccentricity, we also need to understand their distribution in orbital phase, as the energy emitted by the binary system strongly depends on the true anomaly and can vary significantly at different instants $t$ (see \autoref{eq:total_variance}). Consequently, the resulting PSD can fluctuate over time. In the following sections, we estimate the amplitude of these fluctuations.

The one-sided PSD is related to the characteristic strain $h_c$ as $h^2_c(f) = f S(f)$. In \cite{phinney}, it is shown that

\begin{equation}
    h_c^2 (f) = \frac{4G}{\pi c^2 f} \int dz \mathcal{N}(z) \left. \frac{dE_{GW}}{df_r} \right |_{f_r = f(1+z)},
\end{equation}
with $\mathcal{N}(z)dz$ the comoving number density of binaries at redshift $z$. To estimate the GWB spectrum, we need to consider the entire merger history of a binary system. In \cite{Enoki_2007}, the total energy $E_{GW}$ radiated by an eccentric binary system in its entire lifetime is

\begin{equation}
    E_{GW} = \int dt_p L_{GW}(t_p).
\end{equation}

The derivative of $E_{GW}$ with respect to $f_r$ gives

\begin{equation}
\begin{aligned}
    & \frac{dE_{GW}}{df_r} = \int df_p \frac{dt_p}{df_p} \frac{dL_{GW}}{df_r}\\
    & = \int df_p \frac{\tau_{GW} (f_p)}{f_p} \sum_n P_n \delta (f_r - nf_p)\\
    & = \sum_n \frac{\tau_{GW} (f_r/n)}{f_r}  P_n (f_r / n)\\
    & = \frac{\pi}{3G} (G \mathcal{M})^{5/3} (\pi f_r)^{-1/3} \sum_n \left (\frac{2}{n} \right )^{2/3} \frac{g(n, e) f_n(t, e)}{F(e)},
\end{aligned}
\label{eq:dedf}
\end{equation}
where $t_p$ is the time at which the binary has orbital frequency $f_p$ in its rest frame, $dt_p/df_p = \tau_{GW}(f_p) / f_p$ gives the characteristic time that the binary spends at orbital frequency $f_p$ and quantifies the fraction of sources with orbital frequency $f_p$ in a population of binaries. In the last line of \autoref{eq:dedf}, we developed the expression to show the dependence of the energy distribution on chirp mass $\mathcal{M}$, rest frame frequency $f_r$, and eccentricity $e$. In the rest of the article, for the sake of condensed notation, we will express everything as a function of $\tau_{GW}(f)$ and $P_n(f)$. For $\tau_{GW}$, we used

\begin{equation}
    \tau_{GW} = \frac{1}{F(e)}\frac{5}{96} \left ( \frac{c^3}{G\mathcal{M}} \right )^{5/3} (2 \pi f_p)^{-8/3}.
\label{eq:tau_gw}
\end{equation}

In this section, we consider the binaries to be entirely GW-driven, but environmental effects, like stellar hardening, can influence the value of $dt_p/df_p$, especially at low frequencies \cite{Bonetti_2024}. We find the characteristic strain to be

 \begin{equation}
     h_c ^2 (f) = \frac{4G}{\pi c^2 f} \int dz \frac{\mathcal{N}(z)}{(1 + z)^{1/3}} \sum_n \frac{\tau_{GW} (f/n)}{f} P_n (f/n).
\label{eq:enoki_gwb}
 \end{equation}




In the above expression, as in \cite{Enoki_2007}, we consider a population of binaries with characteristic chirp mass $\mathcal{M}$ that have the same initial eccentricity $e_0$ at frequency $f_0$. The frequency and eccentricity of the binaries respectively increase and decrease over time because GWs carry away energy and angular momentum \cite{Loutrel_2018}. In a more realistic formulation, we should consider a distribution of initial eccentricities and chirp masses to account for the astrophysical distribution of the systems \cite{Lamb_2024}.

\begin{figure}
   \centering
   \includegraphics[width=0.5\textwidth]{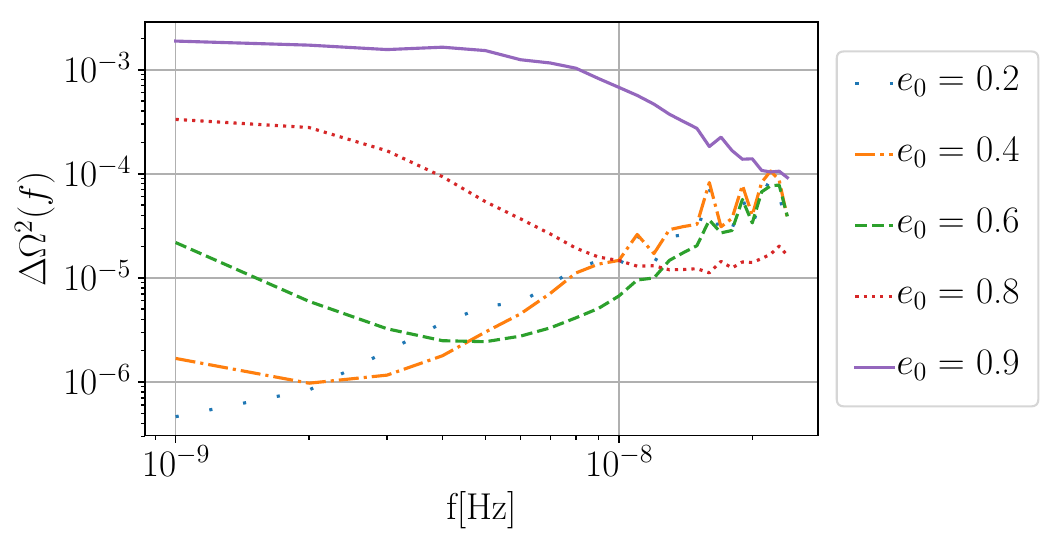}
   \caption{Time-dependent fluctuations $\Delta \Omega^2 (f)$ of the GWB spectrum as a function of frequency $f$ and various initial eccentricities $e_0$ and a population of equal binaries with chirp mass $\mathcal{M}_c$.}
   \label{fig:gwb_fluct}
\end{figure}

To account for individual binary statistics, we have to write the characteristic strain as a sum of individual contributions from each binary at comoving distance $d_M$ \cite{sesana_2008}. Using

\begin{equation}
    h_n ^2 = \frac{4 G}{c^3} \frac{P_n}{n^2 (\pi f_p)^2} \frac{1}{d^2_M},
\label{eq:cgw_h2}
\end{equation}
the characteristic strain of the $n$-th harmonic of a binary \cite{Huerta_2015}, we can write the total characteristic strain $h^2_c$ of a population of $N$ binaries as

\begin{equation}
    h^2_c(f) = \sum_{i=1}^N \sum_n h^2_{n,i} \frac{n f_{p,i}}{\Delta f (1+z_i)} \Theta (f).
\end{equation}

Considering that we have $N(z)$ homogeneously distributed sources between redshift $z$ and $z+dz$, with $p(f_{p,k})$ the fraction of sources with orbital frequency $f_{p,k}$, the total characteristic strain $h_c^2$ is

\begin{equation}
    h^2_c (f) = \sum_z \sum_k \sum_{i=1} ^{N(z)p(f_{p,k})} \sum_n h^2_{n,i} \frac{n f_{p,k}}{\Delta f (1 + z_i)} \Theta (f).
\label{eq:characteristic_strain}
\end{equation}

The term $\Theta (f)$ acts as the binning that keeps only the harmonics with frequency contained between $f$ and $f+\Delta f$ with $\Delta f$ the frequency resolution of the GW detector \cite{Bonetti_2024}. Since we consider a population of equal binaries, the fraction of sources with orbital frequency $f_{p,k}$ is obtained numerically as

\begin{equation}
    p(f_{p,k}) = \frac{\tau_{GW}(f_{p,k}) \Delta \log f_{p,k}}{\sum_k \tau_{GW}(f_{p,k}) \Delta \log f_{p,k}},
\end{equation}
considering that the orbital frequencies of the population of binaries span between $f_{min}$ and $f_{max}$. We will use $f_{min}=10^{-10}$Hz, the orbital frequency at which the population enters the GW emission phase, and $f_{max} \approx 5 \times 10^{-5} $Hz, the frequency of the innermost stable circular orbit for binaries with chirp mass $\mathcal{M} \approx 10^8 M_\odot$, before entering the merging phase \cite{Enoki_2007}.

\subsubsection{Time fluctuations due to eccentricity}

The characteristic strain $h_c$ is related to the noise spectrum as $h^2_c(f) = f S(f)$ \cite{phinney}. Using \autoref{eq:fluct}, we have

\begin{equation}
    \Delta \Omega^2 (f) = \frac{|\langle h^2_c(f)^2 \rangle - \langle h^2_c(f) \rangle^2|}{\langle h^2_c(f) \rangle^2} = \frac{var \{ h^2_c(f) \}}{\langle h^2_c (f)\rangle^2}.
\label{eq:gwb_fluct}
\end{equation}

The characteristic strain $h_c$ is a random quantity since the luminosity $P_n$ is random. The average $\langle x \rangle$ is equivalent to the time average defined in \autoref{sec:power_psi}. Most sources producing a GWB signal in the PTA band are at relatively low redshift $0 < z \leq 1$ \cite{Huerta_2015, Chen_2019}. The variance and mean of $h^2_c$ behave as

\begin{equation}
\begin{aligned}
    var \{ h^2_c(f)\} & \propto \sum_j N(z_j) \frac{d^{-4}_M (z_j)}{(1+z_j)^2} \sum_k p(f_{p,k}) \\
    & \times \sum_{n=n_{low}}^{n_{high}} \sum_{m=n_{low}}^{n_{high}} \left [ cov \{P_n, P_m \} \frac{(n m f^2_{p,k})^{-1}}{\Delta f^2} \right ],
\end{aligned}
\end{equation}

\begin{equation}
\begin{aligned}
    \langle h^2_c(f) \rangle & \propto \sum_j N(z_j) \frac{d^{-2}_M (z_j)}{(1 + z_j)}\sum_k p(f_{p,k})\\
    & \times \sum_{n=n_{low}}^{n_{high}} \left [\langle P_n \rangle \frac{(n f_{p,k})^{-1}}{\Delta f} \right ],
\end{aligned}
\end{equation}
where $n_{low} = f (1+z)/ f_{p,k}$ rounded up to the nearest integer and $n_{high} = (f + \Delta f) (1+z)/ f_{p, k}$ rounded down to the nearest integer. In the expression of the variance, the covariance between $P_n$ and $P_m$ appears when several harmonics of the same binary fall into the same frequency bin $\Delta f$. If no harmonics fall into the bin, the sum is zero.

The sum over $j$ shows the redshift dependent part of the equation that gives the total number of sources contributing to the GWB. For a homogeneous and isotropic Universe, given the density of sources $\mathcal{N}(z) = dN(z)/dV(z)$ and the volume element $dV(z) = 4\pi d^2_M d(d_M)$, the number of sources $N(z)$ at redshift $z$ is $N(z) = \mathcal{N}(z) 4 \pi d^2_M d(d_M)$. The differential of the comoving distance is given by cosmology as $d(d_M)= c dz / H(z)$ with $H(z)$ the Hubble constant at redshift $z$ \cite{hogg2000}. We can then show that the variance and mean of $h^2_c$ in a redshift interval $\Delta z$ scale as

\begin{equation}
    var \{h^2_c (f) \} \sim \frac{\Delta N}{\Delta V} 4\pi\frac{\Delta d}{d^2_M} \sim 8\pi^2 \frac{\Delta N}{\Delta V} \frac{c^2}{H^2(z)}\frac{\Delta z}{\Delta V} \Delta z,
\end{equation}

\begin{equation}
    \langle h^2_c (f) \rangle \sim \frac{\Delta N}{\Delta V} 4\pi  \Delta d \sim 4\pi \frac{\Delta N}{\Delta V} \frac{c}{H(z)} \Delta z,
\end{equation}
so the time fluctuations $\Delta \Omega^2(f)$ in \autoref{eq:gwb_fluct} scale as

\begin{equation}
    \Delta \Omega^2 (f) \sim \frac{1}{\Delta N},
\end{equation}
with $\Delta N$ the total number of sources contained in a volume of Universe $\Delta V$. Similar to Monte Carlo integration, the relative error decreases with the number of samples. For a large number of binaries, the fluctuations become negligible compared to the average spectrum $\langle h_c^2 \rangle$.

In \autoref{fig:gwb_fluct}, we show the expected magnitude of fluctuations for a population of binaries with chirp mass $\mathcal{M} = 10^8 M_\odot$ and various initial eccentricities, focusing on the PTA band between $1$ and $25$ nHz. The average density of sources is assumed to be $\mathcal{N}(z) \approx 2 \times 10^{-3} , \text{Mpc}^{-3}$ (see \cite{Huerta_2015}). The figure highlights the dependence of fluctuations on eccentricity. For a population of highly eccentric binaries, the observed fluctuations are higher at low frequencies. However, we also note that fluctuations increase with frequency, even for populations with low eccentricity. This trend arises because the effective number of binaries contributing to the GWB decreases significantly with frequency. Specifically, because $\Delta \Omega^2$ scales as $1/N$, where $N$ is the number of binaries. The fraction of binaries with orbital frequency $f_p$ is determined by the characteristic time $\tau_{GW}$, as given in \autoref{eq:tau_gw}, where $\tau_{GW} \propto f_p^{-8/3}$. Consequently, fluctuations are expected to grow significantly with frequency, even for populations of binaries with low eccentricity.

We obtained the highest level of fluctuations ($\Delta \Omega^2 \approx 10^{-3}$) for the most eccentric GWB, starting with $e_0 = 0.9$, at the lowest considered frequency $10^{-9}$ Hz. This implies that we need to measure the GWB spectrum with a relative uncertainty of at least $10^{-3}$ to detect its time-dependent features. The fluctuations should be of the order of $\Delta \Omega \approx 1$ to be visible within the current PTA uncertainties on the GWB amplitude estimates \cite{forecasting, falxa_ns} (see \autoref{sec:pta_sensitivity}). As a consequence, it is unlikely that fluctuations from a large population of equal binaries can be detected with current observational abilities; however, newer generations of radio telescopes might offer increased sensitivity to time fluctuations. That said, the astrophysical distribution of the sources in redshift and mass was approximated as uniform in this calculation. In reality, the number of sources contributing to the background at redshift $z$ with chirp mass $\mathcal{M}$ can vary significantly \cite{Bonetti_2024, sesana_2008}. For a realistic population, the variance and mean spectrum in \autoref{eq:gwb_fluct} should be reevaluated for different subsets of the population where $N = \mathcal{N}(z, \mathcal{M}, e)dz d\mathcal{M}dedV$. In this case, fewer nearby of more massive binaries might dominate the total signal, thus generating stronger nonstationarities. In the next section, we account for the effects of the astrophysical distribution of sources using realistic simulations of binary population.

Similarly, one could estimate the frequency-correlated time fluctuations using \autoref{eq:corr_fluct}, yielding

\begin{equation}
\begin{aligned}
    \Delta & \Omega^2 (f, f') = \frac{1}{\langle h^2_c(f)\rangle \langle h^2_c(f')\rangle}\sum_j \sum_k N(z_j) p(f_{p,k}) d^{-4}_M (z_j) \\
    & \times \sum_{n=n_{low}}^{n_{high}} \sum_{m=m_{low}}^{m_{high}} \left [ cov \{P_n, P_m \} \frac{(n m f^2_{p,k})^{-1}}{\Delta f^2 (1+z_j)^2} \right ],
\end{aligned}
\label{eq:gwb_cross_fluct}
\end{equation}
where $m_{low} = f' (1+z)/ f_{p,k}$ rounded up to the nearest integer and $m_{high} = (f' + \Delta f) (1+z)/ f_{p, k}$ rounded down to the nearest integer.

\begin{figure}
   \centering
   \includegraphics[width=0.5\textwidth]{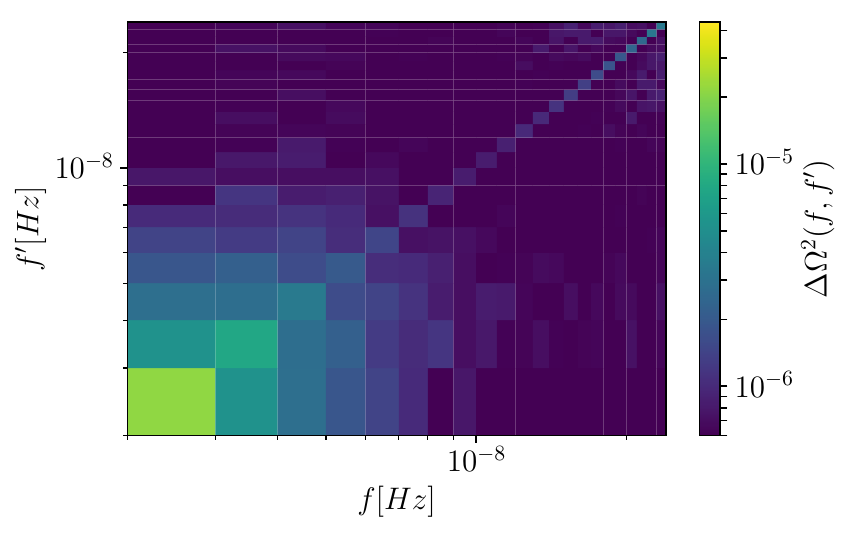}
   \caption{Cross-correlated fluctuations $\Delta \Omega^2 (f, f'$ of the GWB spectrum for frequencies $f$ and $f'$ and initial eccentricity $e_0=0.8$.}
   \label{fig:gwb_cross_fluct}
\end{figure}

\autoref{fig:gwb_cross_fluct} shows the cross-correlated fluctuations for a GWB with initial eccentricity $e_0=0.8$ at $f_0=1$nHz. We see that the correlations between frequencies are highest at low frequencies, where the binaries have the highest eccentricity. All binaries with orbital frequency lower than $10^{-9}$Hz produce a higher concentration of correlated harmonics in the first bins due to the considered PTA resolution of 1 nHz. The presence of correlations in the GWB could be an indicator of nonstationary features induced by eccentric sources \cite{raidal2024}. Furthermore, it is shown in \cite{ns_fluct} that characterizing these correlations provides better constraints and inference of the GWB spectrum properties.

A realistic GWB will have brighter individual sources that stand out from the background spectrum. In the next section, we investigate whether such sources could produce significant fluctuations.

\section{Effect of the astrophysical distribution of binaries}
\label{sec:astro_effect}

In the previous section, we derived the amount of time-dependent fluctuations for a population of equal (relatively light) binaries with chirp mass $\mathcal{M} = 10^8 M_\odot$. A realistic population will have a wide distribution of chirp mass, eccentricity and redshift producing a GWB that can be largely dominated by a small subset of sources \cite{sesana_2008}. Since the fluctuations of the GWB spectrum due to eccentricity is expected to behave as $1/N$ with $N$ the number of sources, in the case where a few bright binaries dominate, we might have very strong fluctuations.

\subsection{Simulated populations}

To simulate an astrophysical population of SMBHBS, we must characterize the number of emitting sources per unit redshift, chirp mass and orbital frequency ${\rm d}^4N/({\rm d}z{\rm d}e{\rm d}\mathcal{M}{\rm dln}f_p)$. The population is obtained by sampling the numerical distribution ${\rm d}^4N/({\rm d}z{\rm d}e{\rm d}\mathcal{M}{\rm dln}f_p)$, which is derived from the observation-based models described in \cite{Sesana_2013} and \cite{Rosado_2015}. Using this method, we draw hundreds of thousands of SMBHBs with astrophysical parameters that are in agreement with current constraints on the galaxy merger rate, the relation between SMBHs and their hosts, the efficiency of SMBH coalescence and the accretion processes following galaxy mergers.

\begin{figure}
   \centering
   \includegraphics[width=0.5\textwidth]{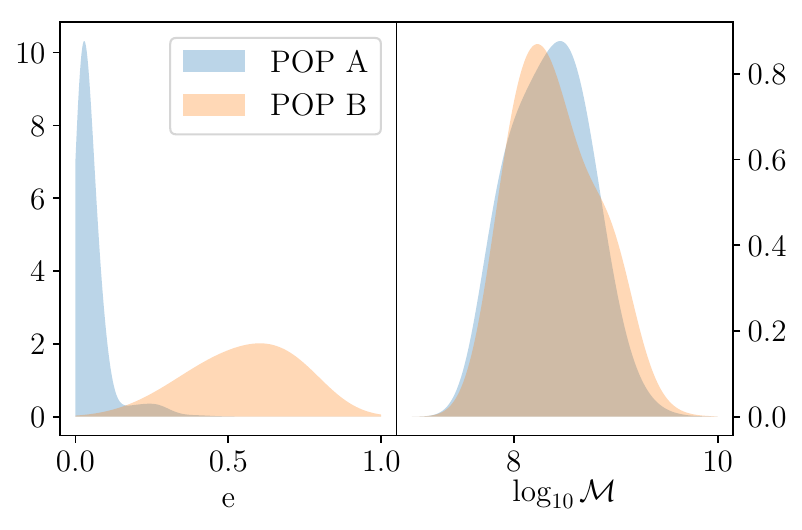}
   \caption{The distribution of eccentricity $e$ and rest frame chirp mass $\mathcal{M}$ (given in solar masses $M_\odot$) for the populations "POP A" and "POP B".}
   \label{fig:pop_props}
\end{figure}

We will compare two populations : "POP A" (low eccentricity) and "POP B" (high eccentricity). In \autoref{fig:pop_props}, we show the distributions of chirp mass $\mathcal{M}$ and eccentricity $e$. Both populations have most binaries distributed with masses between $10^8$ and $10^9 M_\odot$. We see that for "POP B", the distribution of $\mathcal{M}$ is skewed towards higher masses. Considering this small excess at higher masses, combined with high eccentricity, we might already expect stronger fluctuations for this population.

\subsection{Estimating the time fluctuations for realistic populations}

We can compute the mean and variance of the total GWB spectrum by adding the contribution of each source using \autoref{eq:characteristic_strain}, similarly to what was done in the previous section, to the difference that each binary has now a different mass. Using \autoref{eq:gwb_fluct} we can compute the magnitude of time fluctuations $\Delta \Omega^2 (f)$. We do this for 50 different realizations of both populations to get the median and the 1-sigma credible region for $\Delta \Omega^2 (f)$.

\begin{figure}
   \centering
   \includegraphics[width=0.5\textwidth]{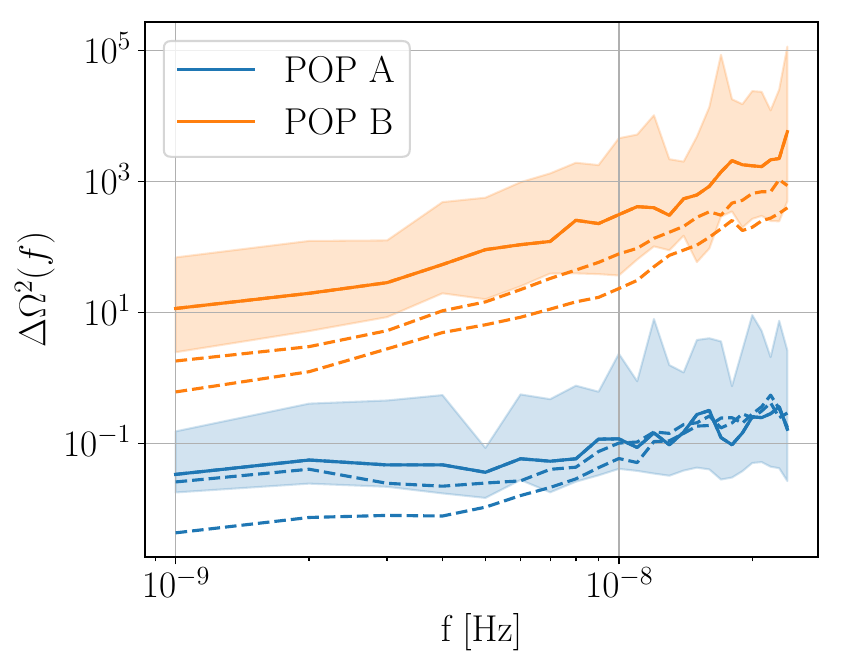}
   \caption{The time fluctuations $\Delta \Omega^2 (f)$ as a function of frequency $f$ for "POP A" and "POP B". The solid lines correspond to median fluctuations for 50 different realizations of the population and the colored areas show the 16\% and 84\% percentiles. The dashed lines are the median fluctuations obtained by removing the 100 and 1000 brightest binaries.}
   \label{fig:realistic_gwb_fluct}
\end{figure}

In \autoref{fig:realistic_gwb_fluct}, we see that the fluctuations are now very high. Indeed, since the populations contain more massive and eccentric binaries, the signal can be dominated by strongly fluctuating sources on top of a fainter population. We clearly see that "POP B" produces much higher values of $\Delta \Omega^2 (f)$ than "POP A" since "POP A" has lower eccentricity. Therefore, there is a clear correlation between the eccentricity of the population and its resulting nonstationarity. Accurately characterizing the nonstationarity of a GWB could allow for a better understanding of the distribution of eccentricities within a population of SMBHBs.

Furthermore, we can also show how the time fluctuations decrease as we remove the brightest binaries of the populations. We estimate the brightness of a source by calculating its SNR $\rho_i^2$ with respect to the total average GWB spectrum $\langle h^2 (f) \rangle$. We have

\begin{equation}
	\rho_i^2 = \sum_n \frac{\langle h^2_{n,i} \rangle}{\langle h^2 (n f'_p) \rangle},
\end{equation}
where $\langle h^2_{n,i} \rangle$ is the average contribution of the $n$-th harmonic of the $i$-th source in the population, and $f'_p = f_p/(1+z)$.

In \autoref{fig:realistic_gwb_fluct}, the dashed lines show the decreasing amplitude of time fluctuations where we removed, respectively in descending order of amplitude, the 100 and 1000 brightest sources. For "POP A," we have relatively low values of $\Delta \Omega^2(f)$ that increase with frequency as the effective number of binaries decreases. Still, $\Delta \Omega^2(f)$ is much higher than what was obtained with the simplified equal binary population of the previous section because there is always a subset of brighter binaries that dominate the signal. By removing the brightest binaries, we show that $\Delta \Omega^2(f)$ declines to lower values. For "POP B," the time fluctuations are much stronger, on the order of $\Delta \Omega^2(f) \approx 10^2$. Since the binaries are highly eccentric, the largest level of fluctuations seems to be pushed to higher frequencies due to the broader harmonic content in the GW spectrum of individual binaries. Overall, accounting for the astrophysical distribution of sources produces nonstationarities that can be detectable by current PTAs, assuming that the pulsar noise models are already well understood \cite{nanograv_noise_budget, wm2}.

\section{Time fluctuations from individual binaries}
\label{sec:time_fluct_cgw}

In this section, we want to estimate the level of fluctuations produced by a single binary standing out of a stationary GWB, given its SNR and eccentricity.

\subsection{Power radiated by individual binaries}

The squared amplitude $h^2_n$ of the $n$-th component of the GW spectrum produced by an eccentric binary is given by \autoref{eq:cgw_h2} where $d_M$ is the comoving distance. The dependence on the distance gives an additional degree of freedom that controls the amplitude of the signal. In the previous section, we have seen that a realistic population of binaries might produce strong fluctuations of the spectrum. In a realistic and heterogeneous realization of the Universe, binaries are spatially distributed as a Poisson process \citep{allen2024sourceanisotropiespulsartiming, Allen_2024}. A single binary can be closer to the observer (or more massive) and appear brighter than the GWB. Considering that $P_n$ is a random variable, we can estimate the variance-covariance between two frequencies as

\begin{equation}
\begin{aligned}
    cov\{h_n^2, h_m^2\} & = \frac{16 G^2}{c^6}\frac{cov\{P_n, P_m\}}{n^2 m^2 (\pi f_p)^4} \frac{1}{d^4_M}.
\end{aligned}
\label{eq:cov_hh}
\end{equation}

The covariance is estimated numerically using \autoref{eq:cov_timefrequency}.

\subsection{Time fluctuations from individual binaries}

Consider a signal $h^2_c$ made up of a stationary background $h^2_{gwb}$ and a single source $h^2$ with orbital frequency $f_p$ and eccentricity $e$:

\begin{equation}
    h^2_c(f, t) = h^2_{gwb}(f) + h^2(f, t).
\end{equation}

The fluctuations will exhibit strong inter-frequency correlations since part of the signal is coming from the same source (see \autoref{fig:covariance_variance}). We calculate the quantity given by \autoref{eq:corr_fluct} 

\begin{equation}
\begin{aligned}
    \Delta \Omega^2 (n f'_p, m f'_p) & = \frac{cov\{h^2_{gwb}(nf'_p) + h_n^2, h^2_{gwb}(m f'_p) + h_m^2\}}{\langle h^2_c(n f'_p, t) \rangle \langle h^2_c(m f'_p, t) \rangle} \\
    & = \frac{cov\{h_n^2, h_m^2\}}{\langle h^2_c(n f'_p, t) \rangle \langle h^2_c(m f'_p, t) \rangle},
\end{aligned}
\end{equation}
where $f_p' = f_p /(1+z)$ and the $h_{gwb}$ terms disappear since the background noise is considered stationary and the only time-dependent components are the $h_n^2$ . Using the above equation, we can calculate the total fluctuations $\Delta \Omega$ of the spectrum by summing over all frequencies:

\begin{equation}
    \Delta \Omega = \sqrt{\sum_{n,m} \Delta \Omega^2 (n f'_p, m f'_p)}.
\end{equation}

We compute the SNR of the source $\rho$ with respect to the stationary background as

\begin{equation}
    \rho^2 = \sum_n \frac{\langle h^2_n \rangle}{h^2_{gwb} (n f'_p)}.
\end{equation}

\begin{figure}
   \centering
   \includegraphics[width=0.5\textwidth]{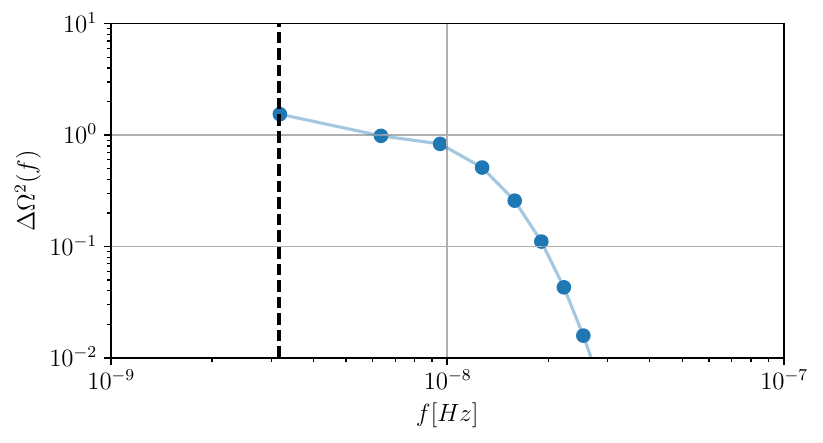}
   \caption{Time-dependent fluctuations of the GWB spectrum as a function of frequency $f$ in the presence of a single source with $\rho = 1.4$, $e=0.5$ and $f_p=5$ nHz. The GWB is considered stationary with a power-law spectrum and similar properties as those found in \cite{nanograv_gwb, wm3}. The vertical dashed line shows the orbital frequency $f_p$ of the individual binary, and the dots show the position of the harmonics.}
   \label{fig:cgw_fluct}
\end{figure}

\begin{figure}
   \centering
   \includegraphics[width=0.5\textwidth]{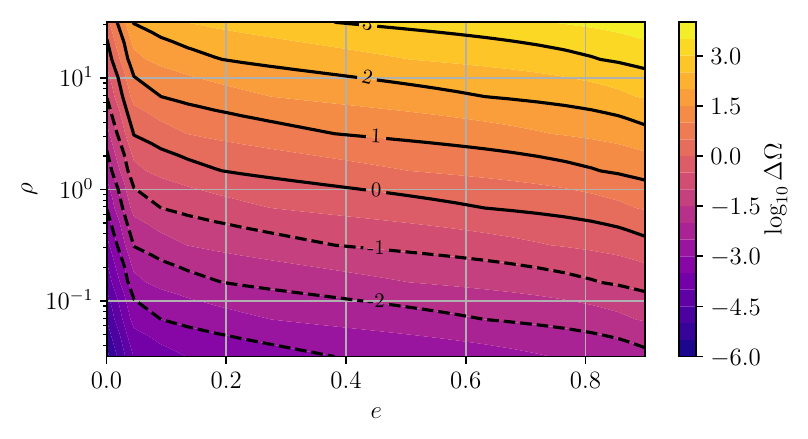}
   \caption{Total time-dependent fluctuations of the GWB spectrum in the presence of a single source for varying $\rho$ and $e$, with $f_p=5$ nHz. The GWB is considered stationary with a power-law spectrum and similar properties as those found in \cite{nanograv_gwb, wm3}. For $e=0$, the fluctuations go to zero because circular binaries do not generate nonstationarities.}
   \label{fig:real_cgw_fluct}
\end{figure}

In \autoref{fig:cgw_fluct}, we show the time fluctuations $\Delta \Omega^2(f)$ produced by a single binary with orbital period $f_p=5\text{nHz}$ and eccentricity $e=0.5$, in the presence of a GWB described as a power law with a spectral index $\gamma \simeq -3.2$ (here converted to characteristic strain $h^2_c \propto f^{2\alpha}$ with $2\alpha = \gamma + 3$), similar to what is found in \cite{wm3, nanograv_gwb}. The SNR of the source with respect to the GWB is $\rho = 1.4$. We see that $\Delta \Omega^2(f)$ is localized around the harmonic spectrum of the source. Even though the SNR is low, the fluctuations it produces are orders of magnitude higher than the ones obtained in \autoref{fig:gwb_fluct} for the whole population of binaries, since it is the brightest among all SMBHBs in this realization of the Universe. We also estimate the total fluctuations for the same source and GWB, but with varying eccentricity and SNR (see \autoref{fig:real_cgw_fluct}). As expected, the fluctuations grow with $\rho$ and $e$ and stay at detectable levels ($\Delta \Omega \approx 1$) for $\rho = 1$. It can be challenging to identify single sources from their deterministic waveform when they have such low SNRs \cite{Becsy_howto, truant2024}. Furthermore, the degeneracy between the GWB and the source makes it difficult to distinguish the two \cite{wm4}. Searching for nonstationarity could offer new ways of identifying each component of the signal.

As we approach $e=0$, $\Delta \Omega$ goes to 0 because circular binaries do not exhibit nonstationary behavior (at least when the frequency evolution of the binary can be neglected). For higher $e$, the fluctuations become more significant. Highly eccentric binaries (parabolic encounters) can produce burst-like signals that are highly nonstationary and are already studied using tailored data analysis methods \cite{Becsy_burst}. We see that for $\rho \approx 1$, $\Delta \Omega$ is close to unity for intermediate eccentricities (as in \autoref{fig:cgw_fluct}). This means that the presence of nonstationary features of a GW signal in a PTA dataset could be compelling evidence of one or a few eccentric single sources.

\section{PTA sensitivity to time fluctuations}
\label{sec:pta_sensitivity}

In order to be detectable, the GWB spectrum variance induced by time fluctuations from eccentric binaries must be greater than the uncertainties on the GWB spectrum amplitude estimates, given a specific PTA configuration (i.e., individual pulsar noise, observation duration, cadence, etc.). A precise estimate of the GWB spectrum can reveal additional properties, such as nonstationarity and non-Gaussianity. In this section, we briefly discuss the implications of noise level for detecting time fluctuations. Recent studies propose more comprehensive statistical analysis methods to detect nonstationary features (see \cite{buscicchio2024, piarulli2024}).

\subsection{Relative uncertainty}

Under the assumption of Gaussian and stationary noise, we can perform a Fisher forecast for the uncertainty on the GWB amplitude estimate following \cite{forecasting}. Considering a GWB with a power-law spectrum $S(f) \propto A^2 f^{-\gamma}$ with fixed spectral index $\gamma$ and amplitude $A$, we define the relative uncertainty

\begin{equation}
    \sigma_S ^2 (f) = \frac{var \{S(f)\}}{S^2_{ML}(f)} = \frac{var\{ A^2 \}}{(A^2_{ML})^2},
\label{eq:relative_uncert}
\end{equation}
where the variance $var\{ A^2 \}$ is given by the 1-$\sigma$ uncertainty from the Fisher estimate of $A^2$, and $A_{ML}^2$ is the best fit maximum likelihood value of the squared amplitude. The relative uncertainty $\sigma_S$ provides a characteristic level of detectability for time fluctuations $\Delta \Omega^2 (f)$. Indeed, if the GWB is nonstationary with a time-varying amplitude $A$ due to the dynamics of its constituents, the evolution of the spectrum will produce a larger variance than in the Gaussian and stationary case, even introducing higher moments (e.g., kurtosis) to the statistical properties of the spectrum \citep{Drasco_2003, PhysRevD.107.063027}. The quantity $\Delta \Omega^2 (f)$ in \autoref{eq:fluct} characterizes the typical variance caused by nonstationary features of the signal PSD. If $\Delta \Omega^2 (f)$ is of the order of (or larger than) $\sigma_S^2 (f)$, the measured variance will be noticeably larger than in the stationary case if the time-dependent features are not properly accounted for \citep{ns_fluct}. In the next subsection, we estimate $\sigma_S$ for different PTA configurations to evaluate their performance in detecting nonstationary fluctuations. A more in-depth analysis would require defining detection statistics, which is beyond the scope of this article.

\subsection{Detectability of time fluctuations}

A PTA dataset is constructed by modeling several physical processes occurring between the emission and reception of pulses at the radio telescope using a timing model \citep{tempo2}. This model predicts the time of arrival (TOA) of pulses with great accuracy. The predicted TOAs are subtracted from the observed TOAs to obtain the timing residuals, which should contain the trace of GW signals as well as unmodeled time-correlated noise components like red noise (RN) and dispersion measure (DM) noise \citep{wm2, nanograv_noise_budget} that can limit sensitivity at low frequencies.

To estimate PTA sensitivity to time fluctuations, we reproduce the noise properties of a given PTA. Based on \cite{forecasting}, we estimate the relative uncertainty $\sigma^2_S$ for three different PTAs:

\begin{itemize}
    \item \textbf{EPTA10}: Based on the EPTA DR2 New dataset used in \cite{wm2, wm3, wm4}, 25 pulsars, 10 years of observation with $\sim$3-day cadence, timing measurement uncertainties $\sim 10^{-6}$s, presence of time-correlated noise RN+DM,
    \item \textbf{IPTA20}: Similar to the EPTA 10-year dataset, 50 pulsars, 20 years of observation with $\sim$3-day cadence, timing measurement uncertainties $\sim$$10^{-6}$s, presence of time-correlated noise RN+DM,
    \item \textbf{SKA10}: A PTA dataset obtained with the next-generation radio telescope Square Kilometer Array (SKA) \citep{Wang_2017}, 30 pulsars, 10 years of observations with $\sim$14-day cadence, timing measurement uncertainties $\sim$$10^{-7}$s, presence of time-correlated noise RN+DM.
\end{itemize}

The algorithm in \cite{forecasting} reconstructs the total sensitivity of a PTA for a given distribution of pulsars and their individual noise properties. This sensitivity is used to estimate the Fisher information matrix for the GWB parameters to forecast uncertainties on the inferred spectrum.

\begin{table}[]
    \centering
    \begin{tabular}{c|c}
        PTA & $\sigma^2_S$ \\
        \hline
        \textbf{EPTA10} & $0.9$ \\
       \textbf{IPTA20} & $9\times 10^{-2}$ \\
        \textbf{SKA10} & $8\times 10^{-3}$
    \end{tabular}
    \caption{Relative uncertainties $\sigma^2 _S$ on the estimate of the GWB amplitude for different PTAs. Nonstationary time fluctuations $\Delta \Omega^2 (f)$ larger than $\sigma^2_S$ are expected to be detectable, as they would introduce significant deviations from the Gaussian and stationary GWB hypothesis.}
    \label{tab:relative_sigma}
\end{table}

As expected, \autoref{tab:relative_sigma} shows that improved sensitivity to fluctuations is achieved with longer, higher-quality datasets and reduced noise (improved GWB SNR). However, caution is needed, as PTA sensitivity remains limited by low-frequency noise sources, such as RN and DM variations \citep{wm2, nanograv_noise_budget}. In this analysis, we considered a fixed spectral index $\gamma$, providing a uniform relative uncertainty $\sigma^2_S$ across all frequencies for a general overview, which is not necessarily true for a realistic PTA. The nonstationary behavior of RN and DM could also introduce bias into the results. Therefore, realistic simulations are essential for accurately assessing the detectability of time-dependent fluctuations.

\section{Conclusion}

We calculated the level of time-dependent fluctuations of the GWB spectrum produced by a population of SMBHBs in the PTA band (1-100 nHz). To do so, we estimated the variance of the GW luminosity of a binary system within one orbital period to characterize the amplitude of GW power emission fluctuations as a function of the binary’s eccentricity. We found that the variance strongly increases with eccentricity and is zero for circular binaries. Calculating this variance requires knowledge of the quadrupole moment.

We showed that for a very large population of homogeneously distributed and equal SMBHBs with chirp mass $\mathcal{M}_c = 10^8 M_\odot$, the amplitude of time-dependent fluctuations of the GWB spectrum is very small compared to the resulting spectrum and is most likely undetectable by current PTA configurations. This is because the fluctuations scale as $1/N$, with $N$ being the total number of binaries in the Universe. Even in the most extreme case where all sources enter the PTA band with an initial eccentricity of 0.9, the fluctuations would be too weak to be measured. However, when accounting for a more realistic astrophysical distribution of sources, we showed that the fluctuations might become very significant and easily detectable if the population is very massive and eccentric. The presence of strong nonstationarities in the GWB spectrum could help us understand the distribution of eccentricity and mass of the SMBHBs making up the population.

Finally, we focused on the scenario where a single binary is brighter than the GWB. We found that even for relatively low SNR (SNR $\approx$ 1) relative to the background noise, the induced time-dependent fluctuations can be significant and detectable. Moreover, they are well-localized around the harmonic frequencies of the GW spectrum of the source. Thus, even for a low SNR binary, detecting significant levels of nonstationarity in the GWB could indicate the presence of one or more bright binary systems in eccentric orbits. That said, this work only briefly explores the detectability of time-dependent fluctuations in the presence of other noise sources, which is left for future investigations. Besides, we derived the GW luminosity fluctuations at leading post-Newtonian order, ignoring the frequency evolution of the binary and higher-order terms. More in-depth analyses with realistic simulations will provide conclusive answers. This work shows the importance of improving our data analysis methods to search for nonstationarities, as they can give us new and unique ways to probe the origins and composition of the GWB signal.

\begin{acknowledgments}
M.F. acknowledges financial support from MUR through the PRIN 2022 project "General Relativistic Astrometry and Pulsar Experiment" (GRAPE, 2022-MYL2X). A.S. acknowledges financial support provided under the European Union’s H2020 ERC Consolidator Grant ``Binary Massive Black Hole Astrophysics'' (B Massive, Grant Agreement: 818691). H.Q-L acknowledges financial support from the French National Research Agency (grant ANR-21-CE31-0026, project MBH\_waves). H.Q-L thanks the Institut Polytechnique de Paris for funding his PhD.
\end{acknowledgments}

\appendix

\section{GW radiation from eccentric sources}

Peter and Mathews \cite{PM} calculated the GW luminosity of eccentric binaries using two different expressions of the quadrupole moment $I^{ij}$ : as a function of the true anomaly and as a Fourier decomposition with respect to the mean anomaly. In this appendix, we calculate the mean and variance of the GW luminosity using both methods and compare the results to verify their validity.

\subsection{True anomaly}

The GW luminosity of a circular binary ($e=0$) is

\begin{equation}
    P_0 = \frac{32}{5} \frac{G^{7/3}}{c^5} (\mathcal{M} 2\pi f_p )^{10/3},
\end{equation}
with $f_p$ the orbital frequency and $\mathcal{M}$ the chirp mass. Following \cite{PM} and using \autoref{eq:quadrupole_formula}, we have the luminosity as a function of true anomaly $\psi$

\begin{equation}
\begin{aligned}
    L_{GW} = \frac{1}{12} \frac{P_0}{(1-e^2)^5} & (1 + e\cos \psi)^4 \\
    & \times [12(1 + \cos \psi)^2 + e^2 \sin ^2 \psi],
\end{aligned}
\end{equation}
that can be averaged over one orbital period. Using

\begin{equation}
    \dot{\psi} = 2\pi f_p \frac{(1 + e\cos \psi)^2}{(1 - e^2)^{3/2}},
\end{equation}
we perform a change of variable and calculate the average as

\begin{equation}
    \langle L_{GW} \rangle = f_p \int_0 ^{2\pi} \frac{d\psi}{\dot{\psi}} L_{GW} = P_0 F(e),
\end{equation}
with the amplification factor given as

\begin{equation}
    F(e) = \frac{\left( 1 + \frac{73}{24} e^2 + \frac{37}{96} e^4\right)}{(1 - e^2)^{7/2}}.
\end{equation}

We can also calculate the variance of the luminosity that quantifies the amount of power fluctuations within one orbital period. We have

\begin{equation}
    \langle L^2_{GW} \rangle = f_p \int _0 ^{2\pi} \frac{d\psi}{\dot{\psi}} L^2_{GW} = P^2_0 G^2(e),
\end{equation}
giving for the variance

\begin{equation}
    \langle L_{GW}^2 \rangle - \langle L_{GW} \rangle^2 = P_0 ^2 [G^2(e) - F^2(e)].
\end{equation}

The recovered function $G^2(e)$ is

\begin{equation}
    G^2(e) = \frac{1 + \frac{271}{12} e^2 + \frac{10155}{128} e^4 + \frac{50966}{768} e^6 + \frac{76735}{6144} e^8 + \frac{1027}{4096} e^{10}}{(1 - e^2)^{17/2}}.
\end{equation}

The factor $G^2(e) - F^2(e)$ quantifies the luminosity fluctuations due to the eccentricity of the binary. In \autoref{fig:sigma_e}, we see that for circular binaries with $e=0$, the fluctuations go to zero.

\subsection{Fourier analysis of Kepler orbit}
\label{app:fourier_kepler}

Using the Fourier analysis of a Keplerian orbit, we can decompose the quadrupole moment in a sum of harmonics $I^{ij} = \sum_n I^{ij}_n$. The second order time derivatives of the $I^{ij}_n$ can be expressed as a combination of coefficients $a_n$, $b_n$ and $c_n$ given in \cite{barack_cutler}

\begin{equation}
\begin{aligned}
a_n & = \frac{1}{2} \left( \ddot{I}^{11}_n - \ddot{I}^{22}_n\right) \\
    & = -n \mathcal{A} [J_{n-2}(ne) - 2eJ_{n-1}(ne) + (2/n)J_n(ne) \\
    & + 2eJ_{n+1}(ne) - J_{n+2}(ne)] \cos [n \Phi(t)], \\
b_n & = \ddot{I}^{12}\\
    & = -n \mathcal{A} (1 - e^2)^{1/2} \\
    & \times [J_{n-2}(ne) - 2J_n(ne) + J_{n+2}(ne)] \sin [n\Phi(t)], \\
c_n & = \frac{1}{2} \left( \ddot{I}^{11}_n + \ddot{I}^{22}_n \right)\\
    & = 2 \mathcal{A} J_n (ne) \cos [n \Phi(t)],
\end{aligned}
\end{equation}
with $\mathcal{A} = [P_0 (5c^5 / 32G)]^{1/2}$ and the $J_n(x)$ are Bessel functions of the first kind. Using \autoref{eq:quadrupole_formula}, we express the luminosity in terms of $a_n$, $b_n$ and $c_n$

\begin{equation}
    L_{GW} = \frac{2G}{5c^5} \left[ (\sum_n \dot{a}_n)^2 + (\sum_n \dot{b}_n)^2 + \frac{1}{3} (\sum_n \dot{c}_n)^2 \right].
\end{equation}

We expand the squared sum and re-express $L_{GW}$ as a sum over harmonics $n$

\begin{equation}
\begin{aligned}
    L_{GW} & = \sum_n \frac{2G}{5c^5} && \bigg[ \dot{a}_n^2 + \dot{b}_n^2 + \frac{1}{3} \dot{c}_n^2 \\
   & && + \sum_{\substack{m \\ m \neq n}} \left(\dot{a}_m \dot{a}_n + \dot{b}_m \dot{b}_n + \frac{1}{3} \dot{c}_m \dot{c}_n \right) \bigg],\\
    & = \sum_n P_n,
\end{aligned}
\label{eq:fourier_basis_lgw}
\end{equation}
where the $P_n$ represent the contribution of each harmonic frequency in $L_{GW}$ at time $t$.

\subsubsection{First moment}

Consider a binary with orbital period $T_p$ and orbital frequency $f_p = 1/T_p$. Following \cite{PM}, we use $\Phi(t) \approx 2\pi f_p (t - t_0)$ and neglect the frequency evolution due the GW radiation of the binary. The brackets $\langle x \rangle$ denote the average over one orbital period. The average luminosity gives

\begin{equation}
\begin{aligned}
    \langle L_{GW} \rangle & = \sum_n \frac{2G}{5c^5} \left \langle \dot{a}_n^2 + \dot{b}_n^2 + \frac{1}{3} \dot{c}_n^2 \right \rangle\\
    & = \sum_n \langle P_n \rangle \\
    & = P_0 \sum_n g(n, e),
\end{aligned}
\end{equation}
where $g(n, e)$ gives the average contribution to the total luminosity of the $n$-th harmonic (the cross terms in $L_{GW}$ with $n \neq m$ give zero when averaged). This result was first derived in \cite{PM}, where $g(n, e)$ is

\begin{equation}
\begin{aligned}
    g(n, e) = & \frac{n^4}{32} \{ [ J_{n-2} (ne) - 2e J_{n-1} (ne) \\
    & + \frac{2}{n} J_n (ne) + 2eJ_{n+1} (ne) - J_{n+2}(ne)]^2 \\
    & + (1-e^2) [ J_{n-2}(ne) - 2J_n(ne) + J_{n+2}(ne)]^2\\
    & + \frac{4}{3n^2}[J_n (ne)]^2 \}.
\end{aligned}
\end{equation}

\subsubsection{Second moment}

For convenience, we write the previously defined $P_n$ as a combination of basis elements $f_n(t, e)$

\begin{equation}
    P_n = \langle P_n \rangle f_n (t, e),
\end{equation}
with $\langle f_n (t, e) \rangle = 1$. The covariance between harmonics $n$ and $m$ is given by

\begin{equation}
\begin{aligned}
    cov \{ P_n, P_m \} & = \langle P_n P_m \rangle - \langle P_n \rangle \langle P_m \rangle\\
    & = P_0 ^2 g(n, e) g(m, e)\\
    & \times [\left \langle f_n (t, e) f_m(t, e) \right \rangle - 1].
\end{aligned}
\end{equation}

In \autoref{fig:covariance_variance}, we see that neighbouring harmonics are very (positively) correlated together. The total variance of $L_{GW}$ is given by

\begin{equation}
    \langle L_{GW}^2 \rangle - \langle L_{GW} \rangle^2 = \sum_{n, m} cov \{ P_n, P_m \}.
\end{equation}

We check the above equality by computing the relative error $\epsilon$ between the analytical total variance $\langle L_{GW}^2 \rangle - \langle L_{GW} \rangle^2$ against the numerically estimated total variance $\sum_{n, m} cov \{ P_n, P_m \}$ as

\begin{equation}
    \epsilon = \frac{|\langle L_{GW}^2 \rangle - \langle L_{GW} \rangle^2 - \sum_{n, m} cov \{ P_n, P_m \}|}{\langle L_{GW}^2 \rangle - \langle L_{GW} \rangle^2}.
\end{equation}

\begin{figure}
   \centering
   \includegraphics[width=0.5\textwidth]{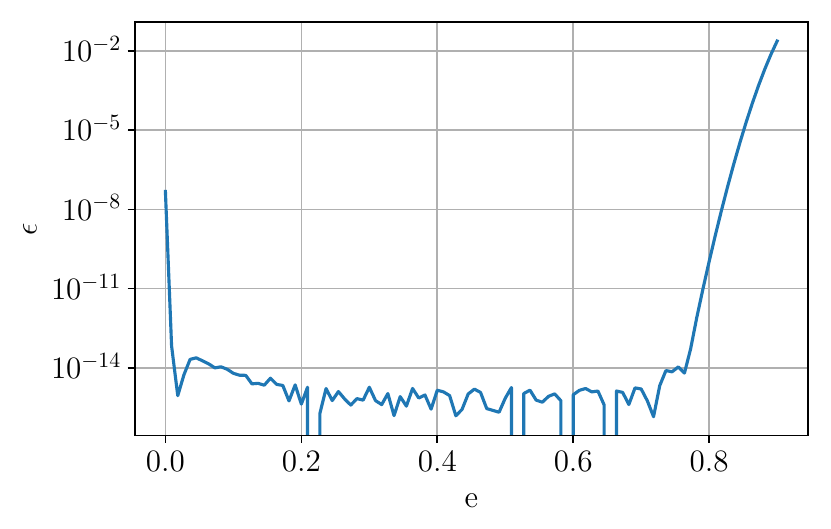}
   \caption{Relative error $\epsilon$ between \autoref{eq:total_variance} and \autoref{eq:cov_timefrequency} as a function of eccentricity $e$. We see that around $e=0.8$, the error goes up. This is because in that example, we used only the first 200 harmonics to estimate the numerical integral in \autoref{eq:cov_timefrequency}. For high $e$, the higher harmonics become significant, so the cutoff at $n=200$ affects our precision.}
\label{fig:tot_cov_error}
\end{figure}

We plot $\epsilon$ as a function of eccentricity in \autoref{fig:tot_cov_error}.

\subsection{Approximated relation $\Delta \Omega = f(e, \rho)$}

Equating \autoref{eq:cov_timefrequency} and \autoref{eq:total_variance} we can roughly approximate the amount of inter-frequency overlap $\left \langle f_n (t, e) f_m(t, e) \right \rangle - 1$ as

\begin{equation}
\begin{aligned}
    & P_0^2 [G^2(e) - F^2(e)] \\
    & = P_0^2 \sum _{n, m} g(n, e) g(m, e) (\left \langle f_n (t, e) f_m(t, e) \right \rangle - 1)\\
    \Leftrightarrow & \sum_{n, m} g(n, e) g(m, e) \left[\left(\frac{G(e)}{F(e)}\right)^2 - 1 \right] \\
    & = \sum_{n, m} g(n, e) g(m, e) (\left \langle f_n (t, e) f_m(t, e) \right \rangle - 1)\\
    \Leftrightarrow & \left[\left(\frac{G(e)}{F(e)}\right)^2 - 1\right] \approx \left \langle f_n (t, e) f_m(t, e) \right \rangle - 1,
\end{aligned}
\end{equation}
where we have used $\sum_{n,m} g(n,e)g(m,e) = F^2(e)$, yielding

\begin{equation}
    cov \{P_n, P_m\} \approx \langle P_n \rangle \langle P_m \rangle \left[\left(\frac{G(e)}{F(e)}\right)^2 - 1 \right].
\end{equation}

Giving an approximated expression for the covariance between $P_n$ and $P_m$. Then, for a single source, we have using \autoref{eq:cgw_h2} 

\begin{equation}
\begin{aligned}
    cov\{h_n^2, h_m^2\} & = 16 \frac{cov\{P_n, P_m\}}{n^2 m^2 f^4_p} \frac{(1+z)^4}{d^4_L}\\
    & \approx 16 \frac{\langle P_n \rangle \langle P_m \rangle}{n^2 m^2 f^4_p} \frac{(1+z)^4}{d^4_L} \left[\left(\frac{G(e)}{F(e)}\right)^2 - 1 \right] \\
    & = \langle h_n^2 \rangle \langle h_m^2 \rangle \left[\left(\frac{G(e)}{F(e)}\right)^2 - 1 \right].
\end{aligned}
\label{eq:approx_cov_hh}
\end{equation}

This expression is very useful since the frequency dependent part factorises from the eccentricity dependent part. Using the above approximated expression for the covariance between $h^2_n$ we have

\begin{equation}
    \Delta \Omega (n f_p, m f_p) \approx \rho^2_n \rho^2_m \left[\left(\frac{G(e)}{F(e)}\right)^2 - 1 \right],
\end{equation}
where we have define the SNR at frequency $nf_p$ as $\rho^2_n = \langle h_n^2 \rangle / \langle h^2_c(nf_p) \rangle$. Then the total fluctuations, summed over all frequencies, give

\begin{equation}
\begin{aligned}
    \Delta \Omega^2 & = \sum_{n, m} \Delta \Omega(n f_p, m f_p) \\
    & = \sum_{n, m} \rho^2_n \rho^2_m \left[\left(\frac{G(e)}{F(e)}\right)^2 - 1 \right],
\end{aligned}
\end{equation}
finally giving

\begin{equation}
    \Delta \Omega^2 = \rho^4 \left[\left(\frac{G(e)}{F(e)}\right)^2 - 1 \right],
\label{eq:total_fluct_approx}
\end{equation}
where $\rho^2 = \sum_n \rho_n^2$ is the total SNR. This expression is convenient to understand the necessary conditions to detect the time fluctuations produced by a single source. In reality, the SNR depends on the frequency of the source $f_p$ and the background noise. Using this approximation, $\Delta \Omega^2$ is only a function of $\rho$, independently of the noise content and $f_p$, giving a practical expression to roughly estimate the level of fluctuations as a function of the SNR and the eccentricity of the source.

\begin{figure}
   \centering
   \includegraphics[width=0.5\textwidth]{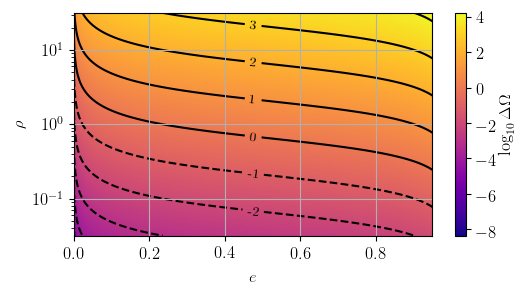}
   \caption{Total time-dependent fluctuations of the GWB spectrum in the presence of a single source for varying $\rho$ and $e$, using the approximation in \autoref{eq:total_fluct_approx}.}
   \label{fig:snr_e_fluct}
\end{figure}

\newpage

\bibliography{ecc_to_ns}

\end{document}